\documentstyle[epsf]{l-aa}
\newcommand{\MBH}{{M}}
\newcommand{\MD}{{M_{_{\rm{D}}}}}
\newcommand{\md}{{m_{_{\rm{D}}}}}
\begin{document}

\thesaurus{02(02.01.2; 02.02.1; 02.09.1)}

\title{On the runaway instability of relativistic tori\thanks{Work
prepared during The Annual Nordic-Trieste Astrophysics Workshop (ICTP,
Trieste 1997)}}

\author{Marek A.\, Abramowicz\inst{1,3,4} \and
        Vladim\'{\i}r Karas\inst{2} \and
        Antonio Lanza\inst{3}}
\offprints{V.~Karas}

\institute{Department of Astronomy and Astrophysics,
G\"oteborg University and Chalmers University of Technology,
S-412\,96~G\"oteborg, Sweden; e-mail:
marek@tfa.fy.chalmers.se \and
Astronomical Institute, Charles University, Faculty of Mathematics
and Physics, V~Hole\v{s}ovi\v{c}k\'ach~2, CZ-180\,00~Praha,
Czech~Republic; e-mail: karas@mbox.cesnet.cz
 \and
Scuola Internazionale Superiore di Studi Avanzati,
Via Beirut 2-4, I-34\,014~Trieste, Italy; e-mail: lanza@sissa.it
 \and
International Centre for Theoretical Physics, Strada Costiera~11, 
 I-34\,014~Trieste, Italy}

\date{Received September 15, 1997; accepted}  

\maketitle
\begin{abstract}

We further investigate the runaway instability of relativistic tori around
black holes assuming a power law distribution of the 
specific angular momentum. We neglect the self-gravity of the torus; thus
the gravitational field of the system is determined by the central
rotating black hole alone (the Kerr geometry). We find that for any positive
(i.e. consistent with the local Rayleigh stability condition) power law
index of the angular momentum distribution, the tori are always runaway
stable for a sufficiently large spin of the black hole. However, before
concluding that some (astrophysically realistic) tori could indeed be
runaway stable, one should include, in full general relativity, the
destabilizing effect of self-gravity. 

\keywords{Accretion: accretion-discs --
          black hole physics --
          instabilities}
\end{abstract}

\section{Introduction} 

Toroidal fluid configurations around black holes, known as thick accretion
disks, have been suggested as models of quasars, other active galactic
nuclei, and some X-ray binaries. Recently they have also assumed an 
important role in cosmological models of
$\gamma$-ray bursts based on the merging of two neutron stars.
A rather simple mathematical class of
such configurations has been introduced by Fishbone \& Moncrief 
(1976), Fishbone (1977), and fully
described analytically by the Warsaw group: Abramowicz, Jaroszy\'nski \&
Sikora (1978), Kozlowski, Jaroszy\'nski \& Abramowicz (1978),
Jaroszy\'nski, Abramowicz \& Paczy\'nski (1980), and Paczy\'nski \& Wiita
(1980). In the last paper, a very practical and accurate Newtonian model
for the gravitational field of a non-rotating black hole, known as the
Paczy\'nski-Wiita potential, was introduced.  

The question of stability of such configurations
attracted considerable attention, because it
was recognized that some types of instability could have very direct, and
quite interesting, astrophysical consequences. Obviously, for the same
reason as in the Newton theory, the essentially {\it local\/} Rayleigh
criterion for dynamical stability with respect to axially symmetric local
perturbations demands that the specific angular momentum $l$ should
increase with the distance $R$ from the axis of rotation (Seguin 1975).
Even the Rayleigh-stable tori with $\mbox{d}l/\mbox{d}R>0$ are 
dynamically unstable in the
presence of a weak magnetic field (Balbus \& Hawley 1991). The
Balbus-Hawley instability does not destroy the large scale structure of
tori, but instead drives a local turbulence which induces viscosity that
is needed for accretion to occur. Much more threatening for the global
structure was the important discovery by Papaloizou \& Pringle (1987) 
that all {\it non-accreting\/} tori are unstable to global
non-axisymmetric perturbations. The consequences of this brilliant work
have been studied by numerous authors, and the present view is that even a
very modest mass loss due to accretion may be sufficient to stabilize the
Papaloizou-Pringle modes (Blaes 1988). Also, it has been
shown that the self-gravity of the disk has a stabilizing effect
(Goodman \& Narayan 1988).

\section{The runaway instability} 

A general feature of fluid distribution around a black hole,
is the existence of a self-crossing equipotential surface, similar
to the Roche lobe in binary systems. The circle at which it crosses
itself is called the cusp. The material inside the Roche lobe
is gravitationally bound and only in such a case 
is equilibrium possible. Those configurations
that just fill their Roche lobes are
marginally bound and thus could be called critical. 
The location of the cusp and the shape of the
relativistic Roche lobe is determined by the combined gravitational
potential of the black hole and the torus, and by the centrifugal
potential due to the disc rotation. 
 
For critical tori, when a small amount of material is accreting 
into the hole, a natural question arises (Abramowicz, Calvani \& Nobili 1983):
is the torus stable with respect to the transfer of mass
through its cusp-like inner edge?  To
answer this question, one must determine the new position of the cusp,
the shape of the Roche lobe, and the new equilibrium for the
torus. If the relativistic
Roche lobe shrinks sufficiently enough, matter which was bounded 
before the mass transfer will
become unbounded, falling catastrophically into
the hole on a dynamical time scale. This is the runaway instability.
A full answer to the question about the conditions
for the runaway instability to occur is far more difficult 
than was originally imagined, and is still unknown today. 
The claim by Abramowicz et
al.\ (1983) that sufficiently massive $l={\rm{const}}$ tori are all runaway
unstable, was based on an approximate model in which the gravitational
field of the central black hole was modeled by the Paczy\'nski-Wiita
potential. The self-gravity of the torus was included using Newtonian
gravity. However, using
the Kerr metric for the black hole, but ignoring the self-gravity
of the torus, Wilson
(1984) demonstrated that the {\em{non self-gravitating}\/} tori do not
suffer from the runaway instability.  The reason is that
accreting matter delivers not only mass but also angular
momentum to the black hole in such a way that the ratio $a/M$ 
(dimensionless specific angular momentum of the black hole)
remains approximatively constant thereby leaving 
unchanged the structure of the space time. By applying another approach (a
Newtonian model of a rotating 
black hole--ring system) Khanna \& Chakrabarti (1992)
confirmed however that self-gravitating rings are runaway unstable. 

It is  obvious that the situation is complex because the self-gravity makes
tori runaway unstable, while rotation of the central black hole
stabilizes them. Thus, both effects are important and neither can be
neglected. To clarify the issue, Nishida et al.\ (1997), using numerically
constructed full general-relativistic models of self-gravitating tori
with $l={\rm{const}}$ found that those that are sufficiently massive
(the ratio $\md=\MD/MBH>0.1$ of the disc mass $\MD$ to the central
black-hole mass $\MBH$) are runaway unstable {\em{always}}. 
These authors pointed out
that the runaway instability introduces an acute difficulty for the
recently considered models for $\gamma$-ray bursts based on the merging
of two neutron stars by making the lifetime of the torus
(which is a product of the merging) too short to provide
enough energy to power the burst.
  
Most recently, however, Daigne \& Mochkovitch (1997) made an important
discovery that a non-zero gradient of the angular momentum distribution
inside the torus has a strong stabilizing effect. They used models of non 
self-gravitating tori and the Paczy\'nski-Wiita potential to describe the black
hole. In this case, after the mass transfer, the angular momentum content
in the innermost part of the resulting configuration is higher and
therefore provides a stronger gravitational barrier 
preventing catastrophic runaway accretion.

To verify that such a stabilizing effect is also present in full general
relativistic regime, one should study the stability of 
equilibrium configurations
constructed self-consistently, in a similar way as in Lanza
(1992) and Nishida \& Eriguchi (1994) but for a non-constant
distribution of angular momentum. 
Constructing general-relativistic equilibrium configurations with
an arbitrary Eulerian rotation law is not
particularly difficult. Explicit analytic models with general 
Eulerian radial
distribution of angular momentum, $l\equiv l(R)$ (for example with a
power-law $l\propto{R^{q}}$) have been constructed by Jaroszy\'nski et al.\
(1978), and later used by Kuwahara (1988), Chakrabarti (1991), and others.
If the equation of state is fixed, each particular model in this class
is determined by the ratio $\md$ of the disc mass $\MD$ to the central
black-hole mass $\MBH$ and by the location of its inner radius.
However, since we want to study the stability of
a torus that is accreting axisymmetrically, it is
necessary to mimic the process by means of a quasi-stationary
sequence of equilibria along which the Lagrangian distribution of
angular momentum per baryon  $j(m)$ is conserved. 
From a numerical point of view 
this is a difficult task. Already in Newtonian theory finding equilibria
with a given Lagrangian angular momentum distribution demands 
a non trivial effort (Ostriker \& Marck 1968, Daigne \& Mochkovitch
1997). The self-consistent
iteration between the Einstein field equations and the equation describing 
the equilibrium of matter is complicated by the fact that the assumed 
distribution of the angular momentum depends on the Lagrangian coordinate
$m$, whereas the field equations are expressed in terms of Eulerian
coordinates (e.g.\ Kozlowski et al.\ 1978). 
(Of course, this problem would disappear if we had spherical
symmetry, since in this case one could easily rewrite Einstein equations
in terms of $m$.) Moreover, in general relativity, 
the surfaces where the
specific angular momentum is constant (von Zeipel surfaces) do
not have a trivial cylindrical structure $R={\rm{const}}$, 
as in Newtonian theory, but must be determined iteratively. 
Their shape is more complicated, given as an
implicit function ${\cal{R}}(R,\theta)={\rm{const}}$, and it is changed
during accretion. This fact is essential for construction of the tori. 
It is convenient therefore to explore such complicated iterative techniques
first in a simpler approximation before solving the full problem.

In this Note we have solved a particular technical problem that is
necessary to obtain the full solution, namely we established a method of
finding non self-gravitating equilibria with a given Lagrangian
distribution of angular momentum in general relativity. 
Apart from a
straightforward replacement of Newtonian formulae by corresponding
relativistic expressions, the major difference and difficulty in studying
relativistic test-fluid tori which we have faced in this work follows from
the fact that the structure of von Zeipel surfaces is not given a-priori
and must be determined iteratively. 

\section{Numerical procedure}

We assume that the spacetime is axially symmetric, stationary,
and described by the Kerr metric $g_{\mu\nu}$ 
in Boyer-Lindquist coordinates $\{t,r,\theta,\phi\}$ 
(geometric units with $c=G=1$ in usual notation). 
The metric is parameterized by 
the black-hole mass $\MBH$ and its specific angular momentum $a/\MBH$.
The torus is described by a relativistic polytrope
which relates the pressure $P$ to the density of material $\rho$: 
$P=\kappa\rho^\gamma$ with $\kappa$ and $\gamma$ constant
(Tooper 1965). The corresponding energy density 
$\epsilon=\rho+P/(\gamma-1)$ is related to the energy per 
unit mass ${\cal{E}}=(P+\epsilon)\rho^{-1}u_t$ and
to Lagrangian angular-momentum density $j={\cal{E}}l$ 
($u_t$ denotes the time-component of the four-velocity of the fluid; 
cf.\ Kozlowski et al.\ 1978).

\begin{figure}
 \epsfxsize=\hsize
 \centering
 \mbox{\epsfbox{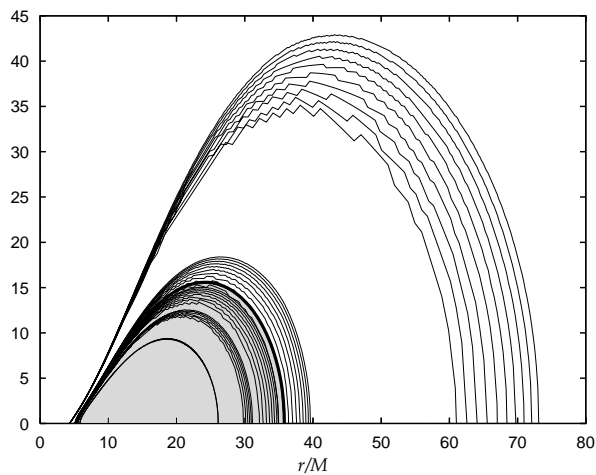}}
 \caption{Illustration of the iterative procedure for constructing a model
 of the torus with a prescribed mass. See text for details.
 \label{fig1}}
\end{figure}

\begin{figure}
 \epsfxsize=\hsize
 \centering
 \mbox{\epsfbox{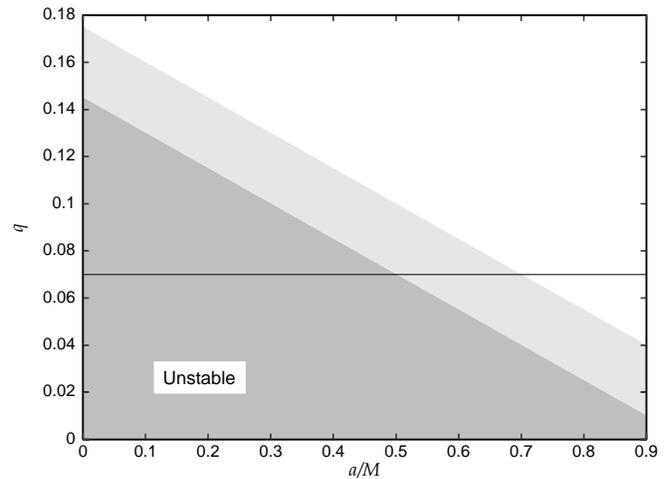}}
 \caption{Shading indicates the region in space of parameters
$(q,a/\MBH)$ in which the tori filling their 
Roche lobes are runaway unstable with respect to small 
overflow. Dark-shaded area
corresponds to the initial mass of the torus $\MD=0.36M_\odot$, 
light-shaded area
refers to $\MD=0.18M_\odot$. Black-hole rotation has apparently
a stabilizing effect. The horizontal line at $q=0.07$ indicates
where the stability boundary is located in the corresponding 
pseudo-Newtonian
analysis of Daigne \& Mochkovitch (1997). See text for details.
 \label{fig2}}
\end{figure}

First, an initial critical configuration is constructed, given 
the parameters $\MD$, $\MBH$, $a$, $\kappa$, and $\gamma$. 
The starting  angular-momentum distribution is specified as a power-law
$l\propto{\cal{R}}^q$ with $q\geq0$ constant. The
model is then  determined by  the above-given parameters,
in particular the inner and the outer edges, 
and the centre of the torus where the pressure is maximum.
The matter distribution inside the disc is calculated and expressed
in terms of a spline-interpolated function of the von Zeipel coordinate, 
$m({\cal{R}})$, where $m({\cal{R}}(r_{\rm{in}}))=0$ 
and normalisation guarantees
$m({\cal{R}}(r_{\rm{out}}))=1$ at the outer edge. Also the Lagrangian 
angular-momentum density
$j$ is expressed as a normalised spline-interpolated function:
${\cal{I}}\equiv{\cal{I}}({\cal{R}})$, $0\leq{\cal{I}}\leq1$. 
This initial configuration
is computed with a pre-determined accuracy of $\MD$.
Figure \ref{fig1} illustrates the above-described iterative procedure. 
 Each bundle of the curves 
 corresponds to a set of surfaces
 with the inner edge fixed and increasing resolution (the effect 
 of improving resolution is clearly visible). Construction of the
 models is terminated when the mass of the torus is determined with
 accuracy better than a
 pre-determined value, typicaly $0.1\%$. Location of the inner edge
 is then changed (another bundle of the curves)  
 and the code iterates again, producing another configuration
 with a different mass of the torus. 
 The whole procedure is repeated (five times in this example) 
 with the inner edge bounded 
 between the marginally bound and the marginally stable orbits.
 Iterations are terminated when
 the torus has a right mass. The final configuration is 
 indicated by shading.

Now we assume that a small amount of material contained near the cusp, between
$r_{\rm{in}}$ and $r_{\rm{in}}+\delta{r_{\rm{in}}}$, is transferred 
throughout the cusp to the black hole. The mass ratio $\md\equiv{\MD/\MBH}$ 
is decreased by $\delta\md$, typicaly of the order of one percent,
while the angular-momentum parameter of the black hole is changed by 
a corresponding value, $\delta\md{l(r_{\rm{in}})}$. The distribution
of $j$ is conserved during the process of accretion:
$j^\prime(m^\prime)=j(m)$ (primes denote new quantities, after the mass
transfer). The next iteration  starts with a guess for the new function
$l_0^\prime=j^\prime/{\cal{E}}$. The inner edge of the new 
critical configuration 
is iterated until a configuration with the required $\md$ is found; at each
iteration $l_0^\prime$ is also updated, so that the final $r^\prime_{\rm{in}}$ 
and $l^\prime({\cal{R}})$ are known only when convergence has been reached.
This updating induces a corresponding change of the von Zeipel
surfaces, ${\cal{R}}(r,\theta)\rightarrow{\cal{R}}^\prime(r,\theta)$
which provides however only a minor modification and is visible
near the inner edge.
At convergence then one can verify whether the inner edge of the new critical
surface has moved.  
If $r_{\rm{in}}+\delta{r_{\rm{in}}}>r^\prime_{\rm{in}}$, then the new
configuration is unstable since the crossing equipotential has shrunk.

\section{Results and Conclusions}
We have adopted the same values of the starting parameters
as Daigne \& Mochkovitch (1997): $\MBH=2.44\,M_\odot$,
$\gamma=4/3$, $\kappa=0.5\times10^{15}$ (cgs), 
corresponding to the models of presumed massive 
tori which are formed in neutron star mergers. The mass of the torus was
chosen to be $\MD=0.36\,M_\odot$ and $\MD=0.18\,M_\odot$, $\delta\md=1\%$. 
We assumed that the torus corotates with the black hole. Figure \ref{fig2}
illustrates the results of the calculation, namely the boundary between
stable and unstable configurations in the plane of parameters
$(q,a/\MBH)$ of the initial configuration. It can be seen that
{\em{}increasing $q$ and increasing $a/\MBH$ both stabilize the 
configuration. This is the main result of our present note.}
The boundary between stability and instability is positioned at roughly 
the same value of $q$ as found by Daigne \& Mochkovitch (1997) 
in their work. The difference is factor of 2 which cannot be considered
too large given the approximate nature of the Paczy\'nski-Wiita
potential and the difference in the iterative procedures. 
Naturally, Daigne \& Mochkovitch 
could not explore the dependence on $a/M$ which does not appear
in the Paczy\'nski-Wiita potential they adopted. 
The horizontal line in Fig. \ref{fig2} indicates
where the boundary is located in the corresponding 
analysis of Daigne \& Mochkovitch with $\MD=0.36\,M_\odot$ when
pseudo-Newtonian radial coordinate is formally identified with
Boyer-Lindquist $r$ in our analysis.

It remains to be discussed how self-gravity of the disc 
affects this criterion of stability of the accretion process,
and what is the interplay between effects of a rotating
black hole and a self-gravitating torus within full general relativity
(work in progress).

\begin{acknowledgements}
We acknowledge clarifying discussions with F.~Daigne.
V.\,K. thanks for hospitality of the International School for
Advanced Studies in Trieste and support from the grant 
GAUK-36/97 of the Charles University in Prague.
\end{acknowledgements}


\begin{thebibliography}{}
\bibitem{}Abramowicz M.~A., Jaroszy\'nski M., Sikora M., 1978, A\&A 63, 221
\bibitem{}Abramowicz M.~A., Calvani M., Nobili L., 1983,
 Nature 302, 597
\bibitem{}Balbus S.~A., Hawley J.~F., 1991, ApJ, 376, 214
\bibitem{}Blaes O., 1987, MNRAS, 227, 975
\bibitem{}Chakrabarti S.~K., 1991, MNRAS 250, 7
\bibitem{}Daigne F., Mochkovitch R., 1997, MNRAS 285, L15
\bibitem{}Fishbone L.~G.,  1977, ApJ, 215, 323
\bibitem{}Fishbone L.~G., Moncrief, V., 1976, ApJ, 207, 962
\bibitem{}Goodman J., Narayan R., 1988, MNRAS, 231, 97
\bibitem{}Jaroszy\'nski M., Abramowicz M.~A., Paczy\'nski B., 1980, 
 Acta Astronomica 30,~1
\bibitem{}Khanna R., Chakrabarti S.~K., 1992, MNRAS 259, 1
\bibitem{}Kozlowski M., Jaroszy\'nski M., Abramowicz M.~A., 1978,  
 A\&A 63, 209
\bibitem{}Kuwahara F., 1988, Progr. Theor. Phys. 80, 449
\bibitem{}Lanza A., 1992, ApJ, 389, 141
\bibitem{}Nishida S., Eriguchi~Y.,  1994, ApJ, 427, 429
\bibitem{}Nishida S., Lanza A., Eriguchi~Y., Abramowicz M.~A., 1997,
 MNRAS 278, L41
\bibitem{}Ostriker J., Mark J.~W., 1968, ApJ, 151, 1075
\bibitem{}Paczy\'nski B., Wiita P., 1980, A\&A, 88, 23
\bibitem{}Papaloizou J.~C.~B, Pringle J.~E, 1987, MNRAS 225, 267
\bibitem{}Seguin F.~H., 1975, ApJ 197, 745
\bibitem{}Tooper R.~F., 1965, ApJ 142, 1541
\bibitem{}Wilson D.~B., 1984, Nature 312, 620
\end{thebibliography}
\end{document}